\documentclass[namedreferences,hyperref,optionalrh]{spr-sola}
\usepackage{graphicx}       
\usepackage{color} 
\ifx\urlurl    \undefined
  \def\urlurl#1{\href{http://#1}{\textsf{#1}}}
\fi

\newcommand{\alfven}{Alfv\'{e}n}
\chardef\us=`\_

\begin{document}

\begin{article}

\begin{opening}

\title{A Statistical Study of Solar White-Light Flares Observed by the \textit{White-light Solar Telescope} of the \textit{Lyman-alpha Solar Telescope} on the \textit{Advanced Space-based Solar Observatory} (ASO-S/LST/WST) at 360 nm}

\author[addressref={aff1,aff2},email={zcjing@pmo.ac.cn}]{\inits{Z. C.}\fnm{Zhichen}~\lnm{Jing}\orcid{0000-0002-8401-9301}}
\author[addressref={aff1,aff2},corref,email={yingli@pmo.ac.cn}]{\inits{Y.}\fnm{Ying}~\lnm{Li}\orcid{0000-0002-8258-4892}}
\author[addressref={aff1,aff2}]{\fnm{Li}~\lnm{Feng}}
\author[addressref={aff1,aff2}]{\fnm{Hui}~\lnm{Li}}
\author[addressref={aff1,aff2}]{\fnm{Yu}~\lnm{Huang}}
\author[addressref={aff1,aff2}]{\fnm{Youping}~\lnm{Li}}
\author[addressref={aff1,aff2}]{\fnm{Yang}~\lnm{Su}}
\author[addressref={aff1,aff2}]{\fnm{Wei}~\lnm{Chen}}
\author[addressref={aff1,aff2}]{\fnm{Jun}~\lnm{Tian}\orcid{0000-0002-1068-4835}}
\author[addressref={aff1,aff2}]{\fnm{Dechao}~\lnm{Song}\orcid{0000-0003-0057-6766}}
\author[addressref={aff1}]{\fnm{Jingwei}~\lnm{Li}}
\author[addressref={aff1}]{\fnm{Jianchao}~\lnm{Xue}}
\author[addressref={aff1}]{\fnm{Jie}~\lnm{Zhao}}
\author[addressref={aff1}]{\fnm{Lei}~\lnm{Lu}}
\author[addressref={aff1}]{\fnm{Beili}~\lnm{Ying}}
\author[addressref={aff1}]{\fnm{Ping}~\lnm{Zhang}}
\author[addressref={aff1,aff2}]{\fnm{Yingna}~\lnm{Su}}
\author[addressref={aff1,aff2}]{\fnm{Qingmin}~\lnm{Zhang}}
\author[addressref={aff1,aff2}]{\fnm{Dong}~\lnm{Li}}
\author[addressref={aff1}]{\fnm{Yunyi}~\lnm{Ge}}
\author[addressref={aff1,aff2}]{\fnm{Shuting}~\lnm{Li}}
\author[addressref={aff1,aff2}]{\fnm{Qiao}~\lnm{Li}}
\author[addressref={aff1,aff2}]{\fnm{Gen}~\lnm{Li}}
\author[addressref={aff1,aff2}]{\fnm{Xiaofeng}~\lnm{Liu}}
\author[addressref={aff1,aff2}]{\fnm{Guanglu}~\lnm{Shi}}
\author[addressref={aff1,aff2}]{\fnm{Jiahui}~\lnm{Shan}}
\author[addressref={aff1,aff2}]{\fnm{Zhengyuan}~\lnm{Tian}}
\author[addressref={aff1,aff2}]{\fnm{Yue}~\lnm{Zhou}}
\author[addressref={aff1,aff3}]{\fnm{Weiqun}~\lnm{Gan}}

\address[id=aff1]{Key Laboratory of Dark Matter and Space Astronomy, Purple Mountain Observatory, Chinese Academy of Sciences, Nanjing 210023, China}
\address[id=aff2]{School of Astronomy and Space Science, University of Science and Technology of China, Hefei 230026, China}
\address[id=aff3]{University of Chinese Academy of Sciences, Nanjing 211135, China}

\runningauthor{Z. Jing et al.}
\runningtitle{A Statistical Study of WLFs Observed by WST}

\begin{abstract}
Solar white-light flares (WLFs) are those accompanied by brightenings in the optical continuum or integrated light. The \textit{White-light Solar Telescope} (WST), as an instrument of the \textit{Lyman-alpha Solar Telescope} (LST) on the \textit{Advanced Space-based Solar Observatory} (ASO-S), provides continuous solar full-disk images at 360 nm, which can be used to study WLFs. We analyze 205 major flares above M1.0 from October 2022 to May 2023 and identify 49 WLFs at 360 nm from WST observations, i.e. with an occurrence rate of 23.9\%. The percentages of WLFs for M1\,--\,M4 (31 out of 180), M5\,--\,M9 (11 out of 18), and above X1 (7 for all) flares are 17.2\%, 61.1\%, and 100\%, respectively, namely the larger the flares, the more likely they are WLFs at 360 nm. We further analyze 39 WLFs among the identified WLFs and investigate their properties such as white-light enhancement, duration, and brightening area. It is found that the relative enhancement of the white-light emission at 360 nm is mostly ($>$90\%) less than 30\% and the mean enhancement is 19.4\%. The WLFs’ duration at 360 nm is mostly ($>$80\%) less than 20 minutes and its mean is 10.3 minutes. The brightening area at 360 nm is mostly ($>$75\%) less than 500 arcsecond$^{2}$ and the median value is 225. We find that there exist good correlations between the white-light enhancement/duration/area and the peak soft X-ray (SXR) flux of the flare, with correlation coefficients of 0.68, 0.58, and 0.80, respectively. In addition, the white-light emission in most WLFs peaks around the same time as the temporal derivative of SXR flux as well as the hard X-ray emission at 20\,--\,50 keV, indicative of Neupert effect. It is also found that the limb WLFs are more likely to have a greater enhancement, which is consistent with numerical simulations. 

\end{abstract}


\end{opening}

\section{Introduction }\label{s-intro} 

Solar flares are a kind of energetic activity with sudden brightening in the solar atmosphere (e.g. \citealp{2011SSRv..159...19F}; \citealp{2011LRSP....8....6S}). White-light flares (WLFs) are the flares with an enhancement in the optical continuum or integrated light (\citealp{1970SoPh...13..471S}; \citealp{1989SoPh..121..261N}; \citealp{2011SSRv..158....5H}). Although the first flare observed in 1859 was a WLF (\citealp{1859MNRAS..20...13C}; \citealp{1859MNRAS..20...15H}), the number of WLFs is very small. Only about 100 WLFs were reported before 2010 \citep{2010ApJ...711..185C}. From 2011 to 2016, about 50 WLFs were observed by the \textit{Solar Optical Telescope} (SOT) on \textit{Hinode} \citep{2017ApJ...850..204W} and no more than 150 WLFs were reported using the data of \textit{Helioseismic and Magnetic Imager} (HMI) on the \textit{Solar Dynamics Observatory} (SDO) (e.g. \citealp{2016ApJ...816....6K}; \citealp{2018ApJ...867..159S}; \citealp{2020ApJ...904...96C}).

Many WLFs are found to have a very strong correlation with the hard X-ray (HXR) and radio emissions in time and space (e.g. \citealp{2006SoPh..234...79H}; \citealp{2011ApJ...739...96K}). This implies that nonthermal electrons are the source of WLFs' energy. However, only the most energetic electrons with more than 900 keV can reach the photosphere \citep{1989SoPh..121..261N} where the white-light (WL) emission is usually emitted (e.g. \citealp{2012SoPh..279..317W}), so it is hard for the nonthermal electrons to heat the photosphere directly. To resolve this problem, much effort has been made to explore the energy transfer and radiative-heating process in the low atmosphere (e.g. \citealp{1986A&A...156...73A}; \citealp{2022A&A...668A..96T}) and many heating sources or mechanisms for WLFs have been proposed, including electron beams (\citealp{1972SoPh...24..414H}; \citealp{1986A&A...168..301A}; \citealp{2020ApJ...891...88W}), proton beams (\citealp{1978SoPh...58..363M}; \citealp{2019PhDT........14P}), chromospheric backwarming (\citealp{1989SoPh..124..303M}; \citealp{2003A&A...403.1151D}), soft X-ray (SXR) irradiation (\citealp{1978SoPh...60..341M}), EUV irradiation (\citealp{1976SvPhU..19..813S}), chromospheric condensations (\citealp{1994ApJ...430..891G}; \citealp{2000A&A...354..691G}), and \alfven\ waves (\citealp{2008ApJ...675.1645F}). In fact, the enhancement of the WL continuum during a flare is likely caused by the joint action of several heating mechanisms (\citealp{2010AN....331..596X}; \citealp{2023ApJ...952L...6S}).

WLFs are observed not only with ground-based telescopes, but also by spacecraft that can provide detailed information. Compared with ground-based telescopes, space telescopes have better and more stable observation conditions. The \textit{Soft X-ray Telescope} (SXT) on \textit{Yohkoh} can provide solar images in the waveband at 430.8 nm with a bandwidth of $\approx$3 nm \citep{1991SoPh..136...37T}. With the SXT data, WLFs were observed from space for the first time \citep{1992PASJ...44L..77H}. The \textit{Transition Region and Coronal Explorer} (TRACE) has a very broad WL channel from 170 nm to $\approx$1000 nm  \citep{2003ApJ...595..483M}. \textit{Hinode}/SOT can provide observations at 430.5 nm with a bandwidth of 0.8 nm \citep{2007SoPh..243....3K}. SDO/HMI gives solar full-disk images at Fe \textsc{i} 617.3 nm \citep{2012SoPh..275..207S}. It should be mentioned that there had been no prior systematic imaging observations in the Balmer continuum (i.e. below the Balmer limit at 364.6 nm) for WLFs, at least for the wavelength range above 300 nm.

The \textit{Advanced Space-based Solar Observatory} (ASO-S; \citealp{2023SoPh..298...68G}) is the first comprehensive solar mission in China, which was launched on 9 October 2022, having a Sun-synchronous orbit with an altitude of $\approx$720 km. The primary aim of ASO-S is to study the magnetic field, solar flares, coronal mass ejections, and their relationships \citep{2019RAA....19..155G}. ASO-S has three payloads: the \textit{Full-disk vector MagnetoGraph} (FMG: \citealp{2019RAA....19..161S}), the \textit{Hard X-ray Imager} (HXI: \citealp{2019RAA....19..160Z,2019RAA....19..163S}), and the \textit{Lyman-alpha Solar Telescope} (LST: \citealp{2019RAA....19..158L}). LST consists of three scientific instruments \citep{2019RAA....19..159C}, a \textit{Solar Disk Imager} (SDI), a \textit{Solar Corona Imager} (SCI), and a \textit{White-light Solar Telescope} (WST). WST can provide continuous full-disk images in the 360$\pm$2 nm waveband \citep{2019RAA....19..162F}, which helps study WLFs at 360 nm, i.e. in the Balmer continuum. Here we still call the flares with 360 nm brightenings WLFs, the same as in \cite{1983SoPh...88..275N}.

To investigate the physical properties of the WL emission at 360 nm for large flares, we study all the M- and X-class flares (i.e. major flares) from 12 October 2022 (when the first M-class flare was observed by WST) to 31 May 2023. 205 M- and X-class flares are collected and 49 of them (23.9\%) are identified as WLFs at 360 nm. We further analyze 39 WLFs among them in detail to investigate the relative enhancement, duration, and brightening area of the WL emission at 360\,nm. In the following, we first introduce the observational data and flare data set in Section \ref{s-instr}. Then we introduce the methods in Section \ref{s-methods}. Our observational results are presented in Section \ref{s-results}. The summary and discussions are given in Section \ref{s-summary}.

\section{Observational Data and Flare Data Set} \label{s-instr}

ASO-S/LST/WST provides the main data in this study. WST works in the 360$\pm$2 nm waveband and has a field of view of 1.2 R$_\odot$. The images taken by WST have a size of 4608$\times$4608 pixels and the pixel scale is $\approx$0.5$^{\prime\prime}$. Note that the spatial resolution of the images (defined as full width at half maximum) is estimated to be about 4$^{\prime\prime}$ and the point spread function is still being under investigation. The cadence of the images is two minutes for a routine mode while it can be as high as one or two seconds in a burst mode. Since the burst mode of WST was under test during the period of our flare data set, the cadence of most flare events is two minutes. The level 2.5 data of WST are used here, which have been corrected for dark current and flat field, as well as made North up and radiometric calibration applied. We further use \textsf{drot\_map.pro} in the SolarSoftWare (SSW) to remove the influence of solar rotation. It should be pointed out that the WST observations were partially affected by Earth eclipses starting around May 2023 and some flares lack available observations unfortunately. 

Considering a strong relationship between the WL and HXR emissions in WLFs, we also use the HXR data from ASO-S/HXI to obtain the peak time of the HXR emission at 20\,--\,50 keV for the WLFs under study. HXI has an energy range of $\approx$10\,--\,300 keV. Its spatial resolution is 3.1$^{\prime\prime}$ and the temporal resolution can be as high as 0.125 seconds. Note that due to the influence of the South Atlantic Anomaly (SAA) or the radiation belt and so on, not all the WLFs at 360 nm have HXR data from HXI.

The SXR 1\,--\,8 \AA\ data from the \textit{Geostationary Operational Environmental Satellite} (GOES) are used in this study as well. The GOES series spacecraft (from GOES-1 to GOES-18) have provided continuous data on solar flares since 1975. The solar X-ray irradiances in 0.5\,--\,4 \AA\ and 1\,--\,8 \AA\ are observed by the \textit{X-ray Sensor} (XRS: \citealp{1996SPIE.2812..344H}) and the latter one is widely used to define the flare magnitude from A- to X-class.

In this work, we collect 205 major flares occurring from 12 October 2022 to 31 May 2023 that have WST observations at 360 nm. There contain 198 M-class and 7 X-class flares with the class from M1.0 to X2.3. Note that there are additional 39 M-class flares observed by GOES but not well observed by WST mainly due to Earth eclipses, which are not analyzed here. From these 205 flares, we identify 49 (23.9\%) WLFs at 360 nm. In our analysis, all the 205 flares collected are used to study the occurrence rate of WLFs at 360 nm as well as compare the SXR duration and spatial distribution of the WLFs with non-WLFs (in Sections \ref{a-occu} and \ref{a-wlfandnwlf}, respectively). Furthermore, for 39 out of the identified 49 WLFs, the physical properties including relative enhancement, duration, brightening area, and total flux of the WL emission at 360 nm are studied (in Sections \ref{sub-param}\,--\,\ref{sub-flux}). In the latter, 10 WLFs near the limb are excluded because their brightenings extend to, or overlap, the solar limb, which leads to an underestimation of the WL area. It should be mentioned that 9 of these excluded 10 WLFs have a longitude of larger than 85$^\circ$.

\section{Methods}
\label{s-methods} 

\subsection{Identification of WLFs at 360 nm}

We use the intensity ratio defined as $(I-I_0)/I_{\mathrm{background}}$ to identify a WLF (see Figure \ref{fig:example}d for an example WLF), where $I$, $I_0$ and $I_{\mathrm{background}}$ are the intensity at 360 nm for a certain pixel in the flare time (i.e. the period between the GOES start and end times of the flare as indicated in Figure \ref{fig:example}f), the mean intensity from the same pixel before the flare onset (30 minutes before the GOES start time, i.e. an interval comparable to the typical duration of a flare), and the mean intensity from a quiet-Sun region before the flare onset (same as the former one), respectively. Note that this ratio is unaffected by the WL brightening region in or outside the sunspot with distinct intensities. It is only affected by a selection of the quiet-Sun region. When selecting the quiet-Sun region, we give preference to those near the flaring active region (see the black box in Figure \ref{fig:example}a) to avoid or reduce the limb darkening effect at 360 nm. Note that the selected quiet-Sun region has a size of 20$^{\prime\prime}\times$20$^{\prime\prime}$, namely $\approx$25 times the spatial resolution of WST. The intensity fluctuation of the quiet-Sun region is checked to be within $\pm$2\% (i.e. the gray line in Figure \ref{fig:example}g) for each WLF. For all the analyzed 39 WLFs, the average level of background fluctuations is $\approx$1.5\%, which is actually similar to the background level of HMI (1.3\%) for the 20 WLFs studied in \cite{2018A&A...613A..69S}. 

Different thresholds of the intensity ratio are used in previous studies for WLFs. For example, a threshold of 5\% was adopted for HMI WLFs at 617.3 nm \citep{2018A&A...613A..69S}. In the present study, we firstly consider the background fluctuations from a quiet-Sun region, i.e. within 2\%. We also consider the fluctuations in the facula region around the flaring active region. In order to make sure that the 360 nm enhancement is caused by the flare itself, we finally adopt 8\% (larger than three times the background fluctuation) as the threshold after some tests to determine a WLF at 360 nm. Note that this threshold is somewhat strict and may have some influence on the WL emission properties, which will be discussed in Section \ref{s-summary}.

\subsection{Calculation of the Brightening Area of WLFs at 360 nm}

When the intensity ratio defined above at a flaring pixel is greater than the threshold of 8\%, the pixel will be counted in the calculation of brightening area for the WLF (see the region marked by a yellow curve in Figures \ref{fig:example}a\,--\,e). We further use the morphological-opening operation for the WL brightening region to remove the noise points \citep{2010SoPh..262..337M}. Note that the brightening area is a function of time and can be written as $S_1$, $S_2$, ..., $S_n$, where $n$ represents the time frame of observations. The final brightening area for a WLF at 360 nm is defined as $S=S_1\bigcup\,S_2\bigcup\,...\bigcup\,S_n$ (see the region marked by an orange curve in Figures \ref{fig:example}d and e). Note that to better understand the WL area and its relationship with some other WL parameters, here $S$ [in units of arcsec$^{2}$] is corrected for projection effect by using the heliocentric angle [$\theta$], i.e. $S=S_0/\cos(\theta)$, where $S_0$ is the originally measured brightening area and $\theta$ is derived from the flare location as listed in Table \ref{tab:all}.

\subsection{Definition of the Duration of WLFs at 360 nm}

Firstly, we use the temporal profile of WL brightening area (i.e. the black curve in Figure \ref{fig:example}g) to determine the start and end times of a WLF at 360 nm. The start time is defined as the first time when the WL brightening area is $>$10 pixels after the flare onset (see the left red vertical line in Figure \ref{fig:example}g), while the end time is defined as the time when the brightening area becomes zero or shows no obvious trend of decline (see the right red vertical line in Figure \ref{fig:example}g). Note that the end time of WL emission can be later than that of the SXR emission. Then, the interval between the start and end times of WL emission is the duration [$\tau$] of WLFs at 360 nm. Considering that the temporal resolution of most WLFs is two minutes, $\tau$ has an uncertainty of four minutes. In addition, the peak time of a WLF at 360 nm is defined as the time when the WL emission reaches its maximum (as marked by the middle red vertical line in Figure \ref{fig:example}g).

\subsection{Definition of the Relative Enhancement of the WL Emission at 360 nm}

The relative enhancement $r$ of the WL emission at 360 nm at a certain pixel is defined as $r=(I-I_0)/I_0$ (e.g. \citealp{2017NatCo...8.2202H}; \citealp{2020ApJ...904...96C}) as shown in Figure \ref{fig:example}e, where $I$ and $I_0$ have the same meanings as mentioned above. Note that $r$ usually has a larger value for the penumbra/umbra pixels compared with the pixels outside the sunspot, as the former pixels have a relatively lower intensity. In the following analysis, we adopt two maximum enhancements for each WLF. One is called ``maximum pixel enhancement" [$r_\mathrm{p}$], representing the maximum enhancement for the brightening pixels during the flare. The other is called ``maximum mean enhancement" [$r_\mathrm{m}$], representing the maximum of the mean enhancement for the WL brightening region during the flare.

\subsection{An Example WLF}

\begin{figure} [htb]
\centerline{\includegraphics[width=1.2\textwidth]{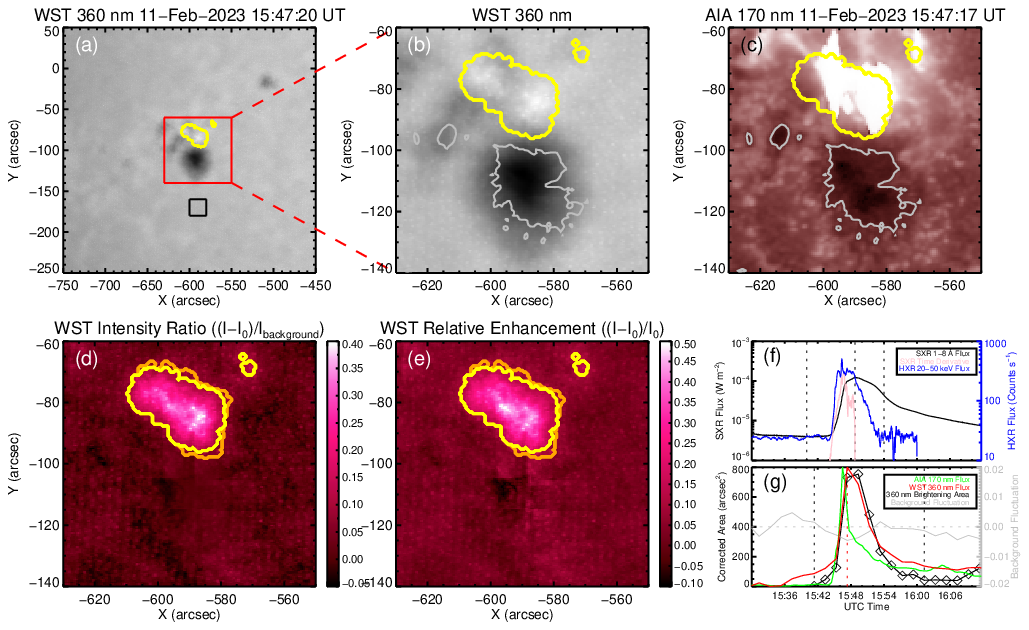}}
\caption{An example of a WLF with a GOES class of X1.1 on 11 February 2023. (\textbf{a}) WST 360 nm image around the flare peak time. The \textit{red box} indicates the flaring region in Panel b. The \textit{black box} denotes a quiet-Sun region that is used to calculate $I_{\mathrm{background}}$. (\textbf{b}) WST image for the flaring region. (\textbf{c}) AIA 170 nm image for the flaring region. The \textit{gray contour} marks the sunspot feature. (\textbf{d}) and (\textbf{e}) Maps of the intensity ratio and relative enhancement of the WL emission at 360 nm. The \textit{yellow curve} (same in Panels a\,--\,e) denotes the brightening area of WL emission at this time, while the \textit{orange curve} (same in Panel d) shows the total WL brightening area of this flare. (\textbf{f}) and (\textbf{g}) Temporal profiles of the SXR 1\,--\,8 \AA\ flux, its temporal derivative, the HXR 20\,--\,50 keV flux from HXI, AIA 170 nm emission (normalized), WST 360 nm emission (normalized), and the brightening area at 360 nm. The \textit{gray curve} in Panel g shows the intensity fluctuation from the quiet-Sun region marked by the \textit{black box} in Panel a. The \textit{three vertical dotted lines} in Panel f denote the GOES start, peak, and end times of the flare. While the \textit{three vertical lines} in Panel g indicate the start, peak, and end times of the WL emission at 360 nm.}
\label{fig:example}
\end{figure}

Figure \ref{fig:example} shows an example WLF with a GOES class of X1.1 on 11 February 2023. From Figures \ref{fig:example}a and b we can see that the identified brightening region at 360 nm (marked by a yellow curve) is near a sunspot, which well matches the brightening kernel at 170 nm (Figure \ref{fig:example}c) observed by the \textit{Atmospheric Imaging Assembly} (AIA) on SDO. The selected quiet-Sun region (denoted by the black box in Figure \ref{fig:example}a) has an intensity fluctuation of less than $\pm$1\%\ as seen from Figure \ref{fig:example}g. For this flare, the intensity-ratio map used to identify the WL brightening (Figure \ref{fig:example}d) looks similar to the enhancement map (Figure \ref{fig:example}e), although the value in the latter one is mostly larger than that from the former one. From Figures \ref{fig:example}f and g, it is seen that the WL emission at 360 nm (integrated over the brightening area, i.e. $S$, as denoted by the orange curve in Figures \ref{fig:example}d and e) is synchronous with the emission at 170 nm roughly, both of which peak around the same time as the HXR 20\,--\,50 keV emission and the temporal derivative of SXR 1\,--\,8 \AA, i.e. the Neupert effect \citep{1968ApJ...153L..59N}. It is also seen that the temporal profile of WL brightening area peaks at a later time than the WL emission.

\section{Results}
\label{s-results} 

\subsection{Occurrence Rate of WLFs at 360 nm}
\label{a-occu}

Based on the above method, we identify 49 WLFs at 360 nm from the collected 205 M- and X-class flares, i.e. with an occurrence rate of 23.9\%. We further check the dependence of the occurrence rate on flare magnitude. As shown in Table \ref{tab:rate}, the percentages of WLFs for M1.0\,--\,M4.9 (31 out of 180), M5.0\,--\,M9.9 (11 out of 18), and above X1 (7 for all) flares are 17.2\%, 61.1\%, and 100\%, respectively. In other words, the larger the flares are, the more likely they are to be WLFs at 360 nm. In particular, all the X-class flares are WLFs at 360 nm in our data set.

\begin{table}[htb]
\caption{Occurrence rate of WLFs at 360 nm.}
\label{tab:rate}
\begin{tabular}{ccccc}                          \hline
Flare class & WLF & non-WLF & Total & Occurrence rate \\
\hline
M1.0\,--\,M4.9 & 31 & 149 & 180 & 17.2\% \\
M5.0\,--\,M9.9 & 11 & 7 & 18 & 61.1\% \\
Above X1.0 & 7 & 0 & 7 & 100\% \\
Total & 49 & 156 & 205 & 23.9\% \\
\hline
\end{tabular}
\end{table}

\subsection{SXR Duration and Spatial Distribution of the WLFs at 360 nm Compared with non-WLFs}
\label{a-wlfandnwlf}

\begin{figure} [htb]
\centering
\includegraphics[width=1.0\textwidth]{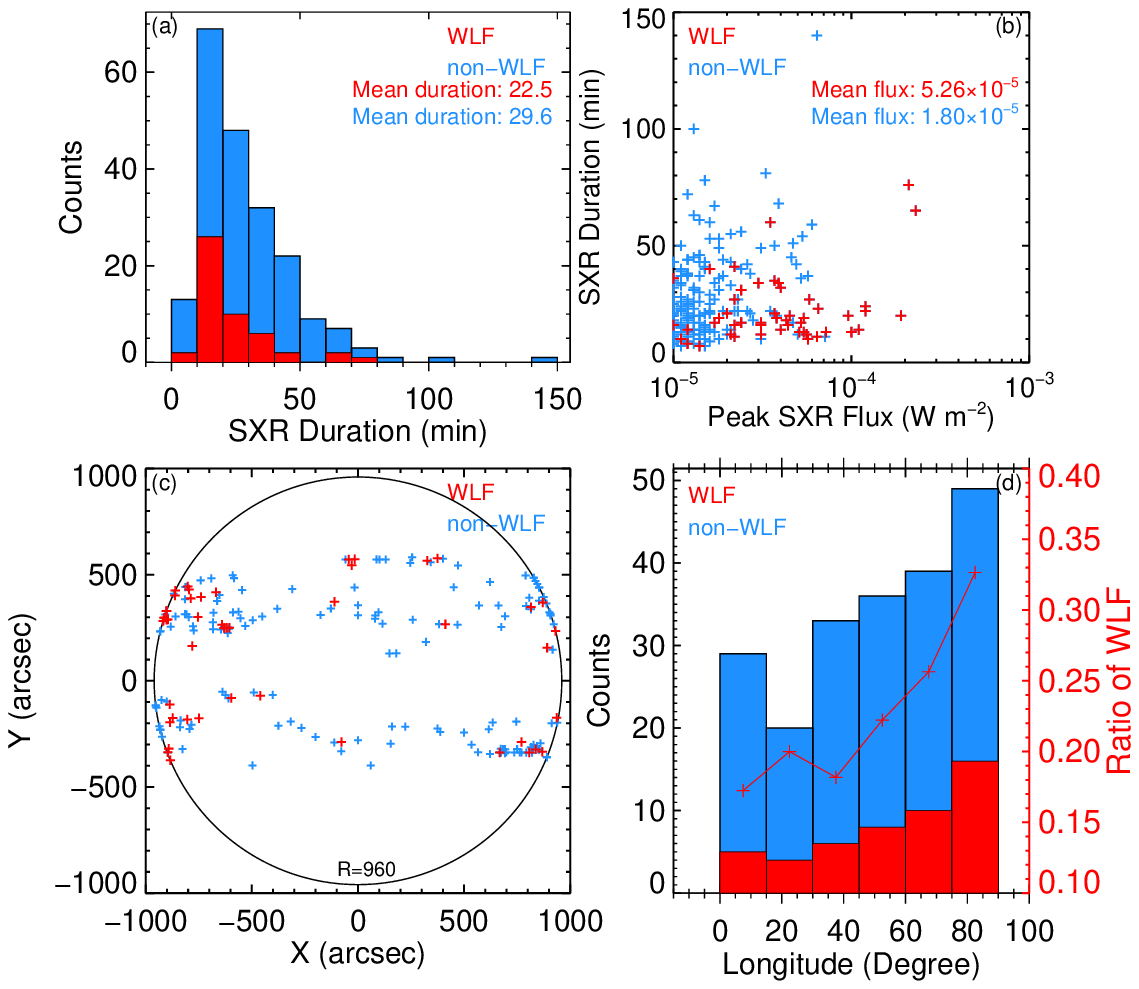}
\caption{Comparisons of WLFs at 360 nm and non-WLFs. (\textbf{a}) Histogram of the SXR duration for all 205 collected major flares. (\textbf{b}) Scatter plot of the SXR duration versus peak SXR flux. (\textbf{c}) Spatial distribution of the WLFs and non-WLFs on the solar disk. (\textbf{d}) Distribution of WLFs and non-WLFs versus solar longitude. The \textit{red curve} plots the ratio of the WLFs. In each panel, the \textit{red and blue colors} represent the WLFs and non-WLFs, respectively.} \label{fig:allfla}
\end{figure}

Figures \ref{fig:allfla}a and b show the histogram of the SXR duration (i.e. the temporal interval between the GOES start and end times) and its scatter plot versus peak SXR flux for all the collected major flares, in which WLFs and non-WLFs are indicated by red and blue colors, respectively. One can see that the SXR durations for all the flares are a few to tens of minutes. Compared with non-WLFs that have a mean SXR duration of 29.6 minutes, the WLFs tend to have a shorter SXR duration, with a mean value of 22.5 minutes. In addition, the WLFs tend to have a larger peak SXR flux (with a mean value of 5.26$\times$10$^{-5}$ W m$^{-2}$) compared with non-WLFs (with a mean value of 1.80$\times$10$^{-5}$ W m$^{-2}$).

Figure \ref{fig:allfla}c shows the spatial distribution on solar disk for all the collected major flares with the WLFs and non-WLFs plotted in red and blue plus symbols, respectively. There are 120 flares on the northern hemisphere and 32 out of them are WLFs at 360 nm, i.e. with a percentage of 26.7\%. For the remaining 85 major flares that are located at the southern hemisphere, there are 17 WLFs at 360 nm among them, i.e. 20\%. Therefore, the spatial distribution of WLFs seems to have no dependence on solar northern or southern hemisphere compared with non-WLFs, at least for our flare data set. However, we notice that there are many of WLFs located near the solar limb. We further investigate the dependence of the WLF's distribution on solar longitude as shown in Figure \ref{fig:allfla}d. It is seen that the number of WLFs exhibits an increasing tendency, say, from several to more than ten, from disk center to limb. Although there are more major flares occurring near the limb in our data set, the percentage of WLFs (i.e. the red curve) still increases from $\approx$17\% to $\approx$33\% over the degree of longitude. This indicates that more WLFs could take place or can be detected near the solar limb.

\subsection{WL Duration, Brightening Area, and Relative Enhancement of WLFs at 360 nm}
\label{sub-param}

In the following, we focus on 39 WLFs (including 6 X- and 33 M-class flares as listed in Table \ref{tab:all}) among the identified 49 WLFs to further study their physical properties. Note that the other 10 WLFs are excluded here due to their brightenings extending to the solar limb, which could lead to an underestimation of the WL area, as mentioned above.

\begin{figure} [htb]
\centering
\includegraphics[width=0.9\textwidth]{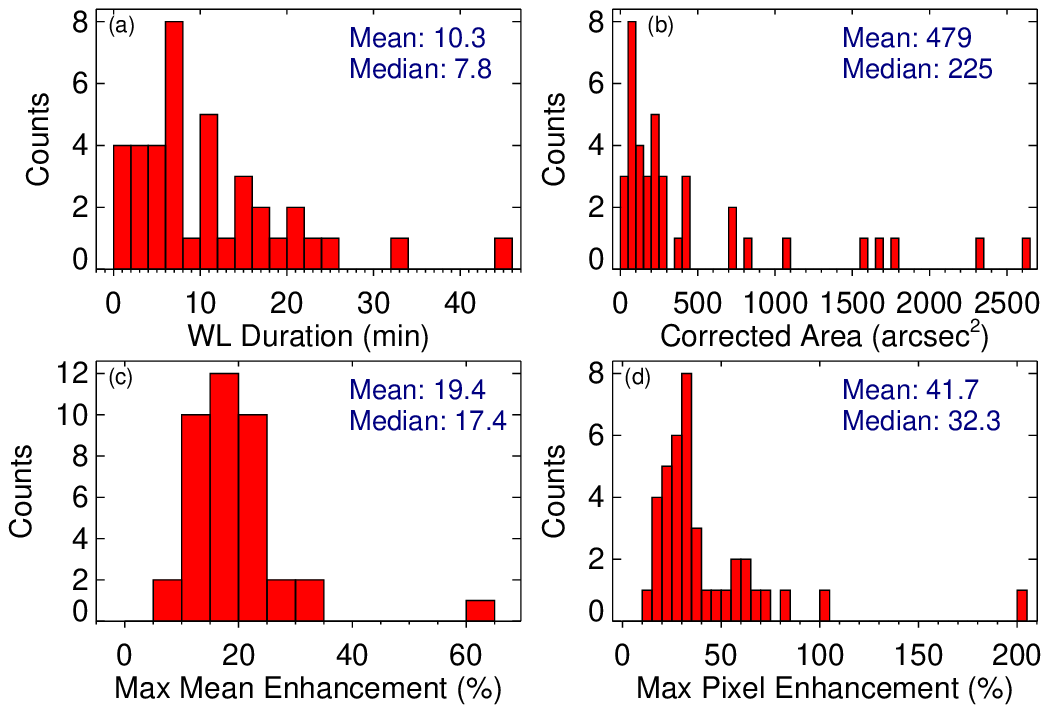}
\caption{WL parameters at 360 nm, including WL duration (\textbf{a}), corrected area (\textbf{b}), $r_\mathrm{m}$ (\textbf{c}), and $r_\mathrm{p}$ (\textbf{d}) for the selected 39 WLFs.}
\label{fig:para}
\end{figure}

Figure \ref{fig:para} shows the histogram of WLFs' parameters including WL duration [$\tau$], brightening area ($S$, corrected for projection effect), maximum mean enhancement [$r_\mathrm{m}$], and maximum pixel enhancement [$r_\mathrm{p}$]. The values of these parameters are also given in Table \ref{tab:all}. From Figure \ref{fig:para}a we can see that most WLFs ($\approx$85\%) have a duration of less than 20 minutes. The longest duration is 44 minutes coming from the X2.3 WLF on 17 February 2023, i.e. the largest flare in our data set. The mean and median durations for all the 39 WLFs are 10.3 and 7.8\,minutes, respectively. From Figure \ref{fig:para}b it is seen that most WLFs ($\approx$75\%) have a brightening area of smaller than 500 arcsec$^2$ and the mean/median area of all WLFs is 479/225 arcsec$^2$. For $r_\mathrm{m}$ and $r_\mathrm{p}$ shown in Figures \ref{fig:para}c and d, one can see that most of WLFs have a $r_\mathrm{m}$ of less than 40\% and a $r_\mathrm{p}$ of less than 75\%. The mean $r_\mathrm{m}$ and $r_\mathrm{p}$ of all WLFs are 19.4\% and 41.7\%, respectively, and the medians are 17.4\% and 32.3\%, respectively. Note that the X2.1 WLF on 3 March 2023 (the second largest flare in our data set) near the limb has the largest enhancement, with a $r_\mathrm{p}$ of 203\% and a $r_\mathrm{m}$ of 62.1\%.

\begin{figure} [htb]
\centering
\includegraphics[width=0.95\textwidth]{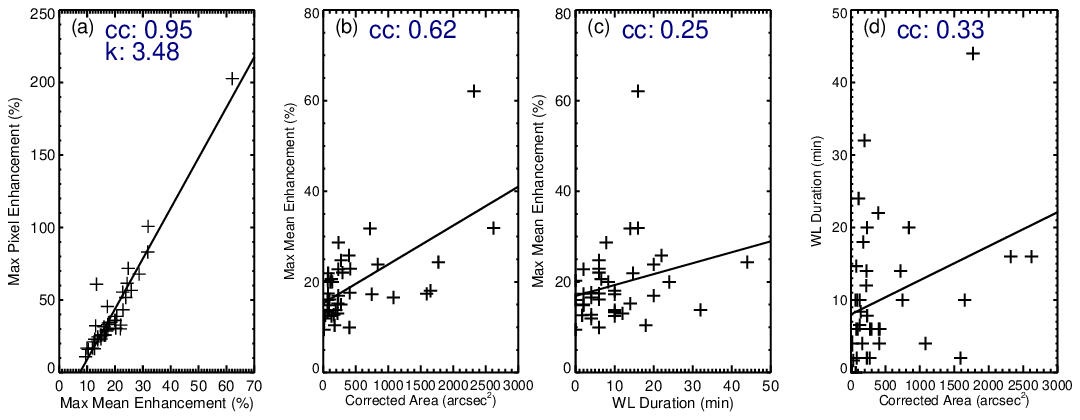}
\caption{WL parameters and the linear fit for their relationships.}
\label{fig:relation}
\end{figure}

The relationships among the above WL parameters are further studied by calculating their correlation coefficients (cc), as shown in Figure \ref{fig:relation}. We can see that $r_\mathrm{m}$ and $r_\mathrm{p}$ of all the 39 WLFs have a strong positive correlation with cc$=$0.95 (see the scatter plot in Figure \ref{fig:relation}a). $r_\mathrm{m}$ also has a good relationship with the corrected area (cc$=$0.62, Figure \ref{fig:relation}b). While $r_\mathrm{m}$ has no obvious relationship with the WL duration (cc$=$0.25, Figure \ref{fig:relation}c). In addition, the WL duration has a poor correlation with the corrected area (cc$=$0.33, Figure \ref{fig:relation}d).

\begin{figure} [htb]
\centering
\includegraphics[width=0.9\textwidth]{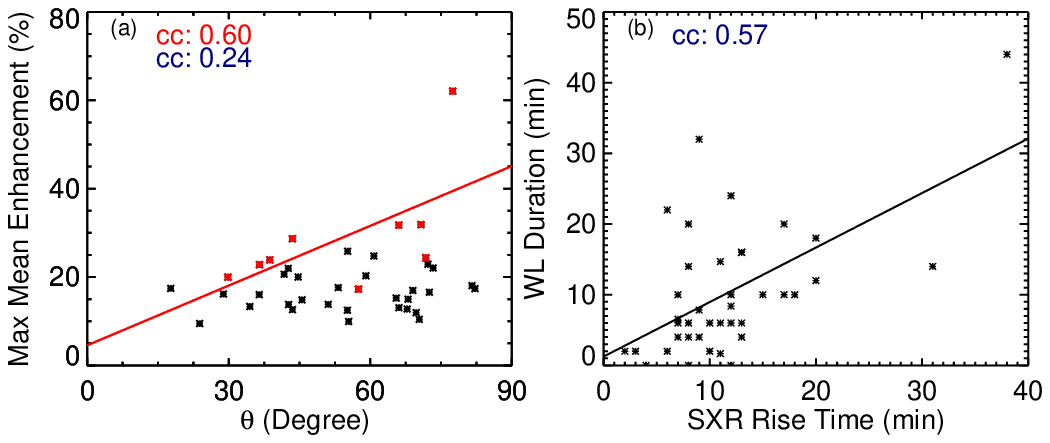}
\caption{(\textbf{a}) $r_\mathrm{m}$ versus heliocentric angle. The \textit{red symbols} denote the WLFs above M6 and the \textit{red line} is their linear fit. (\textbf{b}) WL duration versus SXR rise time with a linear fit.} \label{fig:other}
\end{figure}

Figure \ref{fig:other}a shows the center-to-limb distribution (in terms of the heliocentric angle [$\theta$] from 0 to 90$^\circ$) of $r_\mathrm{m}$ for all the 39 WLFs. It is seen that as $\theta$ increases, $r_\mathrm{m}$ reaches a higher value, though a small $r_\mathrm{m}$ can also show up near the limb. The cc between $r_\mathrm{m}$ and $\theta$ for all the 39 WLFs is 0.24, i.e. showing a weak relevance. However, if we only choose WLFs above M6.0 (see the red star symbols), there exists a much stronger correlation (cc$=$0.60) between $r_\mathrm{m}$ and $\theta$. This suggests that $r_\mathrm{m}$ has a center-to-limb variation especially for large WLFs, which will be discussed in Section \ref{s-summary}.

In addition, we check the relationship between WL duration and SXR rise time (i.e. the time period from flare onset to peak) for the 39 WLFs, as shown in Figure \ref{fig:other}b. There shows a good positive correlation (with cc$=$0.57) between these two parameters. Note that the WL duration has a comparable value with the SXR rise time. This implies that the WL emission is mainly caused by the heating process as indicated by the SXR rise time of a flare.

\begin{figure} [htb]
\centering
\includegraphics[width=1.0\textwidth]{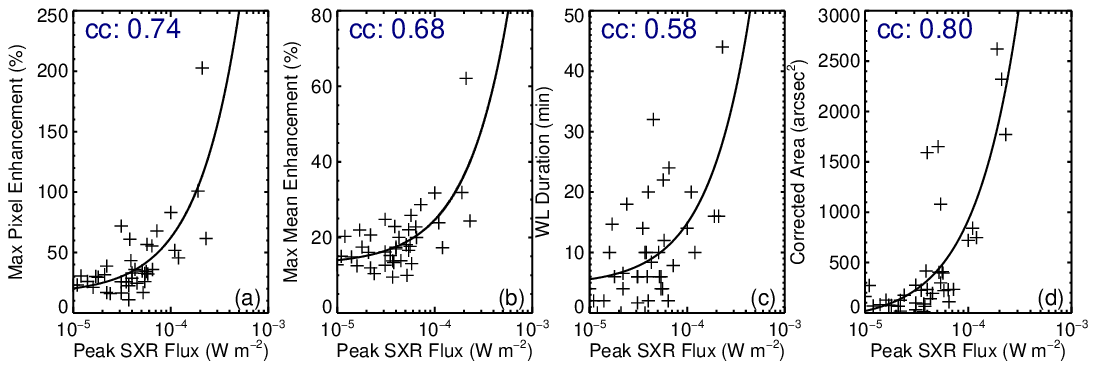}
\caption{WL parameters versus peak SXR flux with a linear fit.}
\label{fig:flux}
\end{figure}

The dependencies of WL parameters at 360 nm on flare magnitude or peak SXR flux are shown in Figure \ref{fig:flux}. We can see that all the parameters, $r_\mathrm{p}$, $r_\mathrm{m}$, $\tau$, and $S$, have a good relationship with the peak SXR flux, with cc of 0.74, 0.68, 0.58 and 0.80, respectively. This is reasonable, since a larger flare tends to have a more energy, which can lead to a stronger WL brightening.

\subsection{Relationships of the WL Peak Time with SXR and HXR Peak Times}
\label{sub-hxr}

\begin{figure} [htb]
\centering
\includegraphics[width=1.0\textwidth]{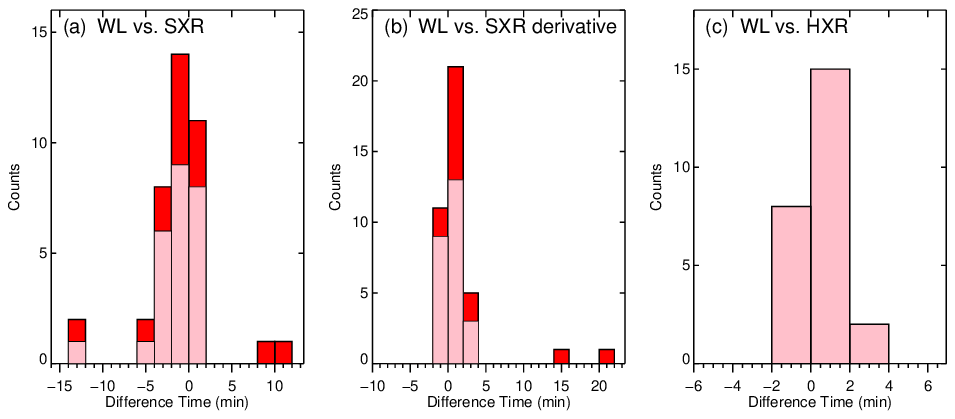}
\caption{Differences between WL and SXR peak times (\textbf{a}), between WL peak time and peak time of SXR temporal derivative (\textbf{b}), and between WL and HXR peak times (\textbf{c}). In each panel, the \textit{pink and red colors} represent the WLFs with and without HXI data available, respectively.} \label{fig:peakdiff}
\end{figure}

Comparing the WL peak time to the SXR or HXR peak time can help us understand the physical origin of WL emission in flares. Figure \ref{fig:peakdiff}a plots the histogram of the time difference between WL and SXR peak times for all the 39 WLFs. Note that the pink and red colors refer to the WLFs with and without available HXR 20\,--\,50 keV observations from ASO-S/HXI, respectively. It is seen that two thirds (26) of the WLFs have an earlier WL peak time compared to the SXR peak time. Considering an uncertainty of $\pm$2 minutes for the WL peak time, there are still 12 WLFs (about one third) having a time difference of less than $-$2 minutes, while only two WLFs have that of more than two minutes. Therefore, the WL emission tends to reach its peak earlier than the SXR emission. This also implies that the WL emission is mainly associated with the heating process of the flare, i.e. similar to the indications of the relationship between the WL duration and SXR rise time as mentioned above. Note that there are two WLFs (an M4.5 and an M1.6) whose WL emissions peak much later ($\approx$ten minutes) than the SXR emission. These gradual-phase WL emissions might be more related to the cooling process of the flare.

We further check the difference between the WL peak time and the peak time of SXR temporal derivative as shown in Figure \ref{fig:peakdiff}b. We can see that most of the WLFs (95\%) have a peak time of WL roughly the same as that of the SXR temporal derivative, with the time difference of $\pm$4 minutes, including all the 25 WLFs with the HXR data available (see the pink color). This indicates that the WL emission at 360 nm exhibits the Neupert effect \citep{1968ApJ...153L..59N} in principle. The two exceptional WLFs are those having a later WL peak time as mentioned before. They also have a later WL peak time than the one of SXR temporal derivative.

For the 25 WLFs with HXR data available, we further plot the histogram of the difference between WL and HXR peak times in Figure \ref{fig:peakdiff}c. One can see that all these WLFs have a similar peak time for the WL and HXR emissions within a range of $\pm$4 minutes and particularly most (92\%) are in the range of $\pm$2 minutes. This confirms that the WL emission at 360 nm basically follows the Neupert effect and thus implies that the 360 nm emission is closely related to nonthermal electron-beam heating, at least for most WLFs under study.

\subsection{Total Flux of WLFs at 360 nm}
\label{sub-flux}

\begin{figure} [htb]
\centering
\includegraphics[width=1.0\textwidth]{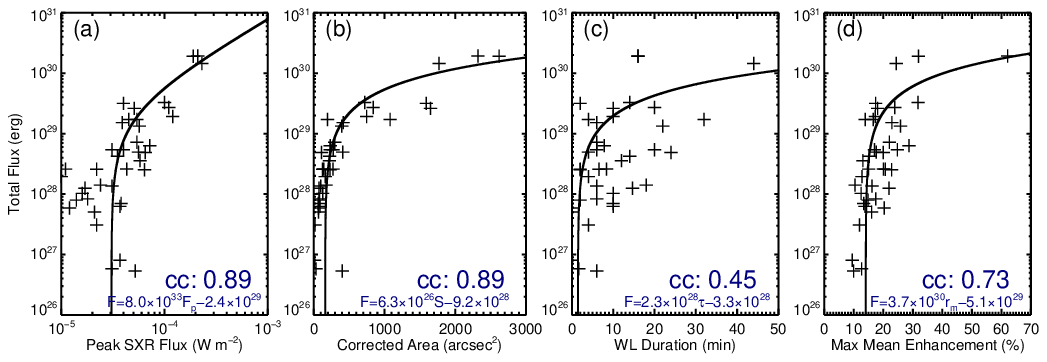}
\caption{Total flux of the WLFs versus peak SXR flux (\textbf{a}), corrected area (\textbf{b}), WL duration (\textbf{c}), and $r_\mathrm{m}$ (\textbf{d}) with a linear fit. The fit expressions are given in the panels, where $F_\mathrm{p}$ in Panel a represents the peak SXR flux of WLFs.} \label{fig:totalinten}
\end{figure}

To describe the absolute enhancement of WLFs at 360 nm, we define the total flux (or total energy) as: $F=\sum\limits_t\sum\limits_{S_\mathrm{t}}(I-I_0)$ \citep{2020ApJ...904...96C}, i.e. integrating the enhanced WL intensity over corrected area [$S_\mathrm{t}$] and time [$t$]. Note that here the WL intensities of $I$ and $I_0$ have been calibrated into a physical unit of erg s$^{-1}$ cm$^{-2}$ sr$^{-1}$. In addition, the limb-darkening effect at 360 nm has been corrected by using an empirical formula of $\frac{I(\mu)}{I(\mu=1)}={u}_{\lambda}+(1-{u}_{\lambda})\mu+{v}_{\lambda}\mu\ln{\frac{2\mu}{\mu+1}}+{w}_{\lambda}\mu(\mu\ln{\frac{\mu+1}{\mu}}-\ln2)$, where ${u}_{\lambda}$, ${v}_{\lambda}$, and ${w}_{\lambda}$ are 0.2887, 2.2194, and $-$5.6933, respectively \citep{1991SvA....35..441M}. This formula can fit the WST data quite well when $\mu>0.11$. It should be mentioned that all the analyzed 39 WLFs have $\mu\ge0.13$. The final $F$ is listed in Table \ref{tab:all}. It is seen that the total flux at 360 nm with a bandwidth of $\pm$2 nm for all the 39 WLFs is about 10$^{26}$\,--\,10$^{30}$ erg. The relationships between the total flux and peak SXR flux (or flare magnitude), brightening area, WL duration, and maximum mean relative enhancement are shown in Figure \ref{fig:totalinten}. We can see that the total flux has a good positive correlation with these four parameters in general, with cc of 0.89, 0.89, 0.45, and 0.73, respectively. This result is reasonable and can be predicted. Here we also give the expression of $F$ based on the linear fitting (see each of the panels), which could be used to estimate the radiated energy at 360$\pm$2 nm via the peak SXR flux only and further estimate the other WL parameters at 360 nm roughly.

\section{Summary and Discussions}
\label{s-summary} 

In this article, we present a statistical study on the WLFs at 360 nm observed by ASO-S/WST from October 2022 to May 2023. The occurrence rate of WLFs and comparisons between WLFs and non-WLFs are provided. The WLFs' parameters including relative enhancement ($r_\mathrm{m}$ and $r_\mathrm{p}$), WL duration, brightening area, and total flux are given, together with their mutual relationships. The WL peak time is also compared to the SXR and HXR peak times, which is helpful to understand the physical origin of WL emission. Our results are summarized below.

\begin{enumerate}
\item The occurrence rate of WLFs at 360 nm is 23.9\% for major flares above M1.0, which increases over the flare magnitude. It should be mentioned that all the seven X-class flares in our data set are WLFs at 360 nm.

\item Compared with non-WLFs, the WLFs at 360 nm tend to have a shorter SXR duration and a larger peak SXR flux. The spatial distribution of WLFs seems to have no dependence on solar northern or southern hemisphere, at least for our flare data set. However, the number of WLFs increases from disk center to limb and the relative enhancement of WL emission can be larger for the WLFs near the limb.

\item The mean values of WL duration, brightening area, $r_\mathrm{m}$, and $r_\mathrm{p}$ of WLFs at 360 nm are 10.3 minutes, 479 arcsec$^2$, 19.4\%, and 41.7\%, respectively. $r_\mathrm{m}$ is found to have a strong relevance with $r_\mathrm{p}$ and some positive relevance with the area. However, it has no obvious relevance with the duration. In addition, there is no obvious correlation between the duration and area.

\item All the WL parameters, including the duration, area, $r_\mathrm{p}$, $r_\mathrm{m}$ and total flux, show a good positive relationship with the peak SXR flux.

\item The WL emission at 360 nm reaches its peak around the same time as the temporal derivative of SXR flux as well as the HXR 20\,--\,50 keV emission for most WLFs. This indicates that the WL emission basically follows the Neupert effect. It should be noted that there are two WLFs showing a notable gradual-phase emission at 360 nm. 

\item The total flux at 360 nm for the 39 WLFs under study is about 10$^{26}$\,--\,10$^{30}$ erg. It has a good positive correlation with the peak SXR flux, WL enhancement, and brightening area.
\end{enumerate}

Our study presents a continuous observation of major flares at 360 nm for more than half a year from ASO-S/WST as well as provides the occurrence rate of WLFs at 360 nm for the first time. Although there are some other telescopes observing the Sun in the 360 nm waveband, such as the \textit{Optical and Near-infrared Solar Eruption Tracer} (ONSET: \citealp{2013RAA....13.1509F}), the advantages of WST are the continuity of its observation in a seeing-free condition from space and its full-disk field of view being beneficial for flare studies. Nearly 50 WLFs have been identified in the period of less than one year, which is roughly equivalent to the number of WLFs observed by HMI at 617 nm for about 2\,--\,3 years. Note that our threshold for identifying a WLFs at 360 nm is relatively high and the burst mode of WST for a very high cadence was still under test. Therefore, there should be more WLFs detected by WST during this period. Here we should also mention that WST and HMI work in different wavebands and the occurrence rate of WLFs in these two wavebands can be different.

Based on some statistical studies on HMI WLFs at 617 nm in the literature, here we can roughly compare the properties of WL emissions between WST 360 nm and HMI 617 nm. For instance, the mean enhancement of the emission at 617 nm is $\approx$15\% (e.g. \citealp{2018A&A...613A..69S}, \citealp{2020ApJ...904...96C}), and $r_\mathrm{m}$ of the emission at 360 nm is $\approx$19\%. Considering that a higher threshold of intensity ratio for WLFs, i.e. 8\% in the present study, would overestimate a $r_\mathrm{m}$, we may say that the mean enhancement at 360 nm is comparable to that at 617 nm in WLFs. In addition, the duration and area of WLFs at 617 nm have median values of $\approx$5 minutes and $<$50 arcsec$^2$ (without a correction of project effect; \citealp{2018A&A...613A..69S}), respectively, both of which are smaller than those of WLFs at 360 nm. Note that a higher threshold for WLFs here would underestimate the brightening area and duration of the emission at 360 nm. A detailed comparison between the emissions at WST 360 nm (in the Balmer continuum) and HMI 617 nm (in the Paschen continuum) deserves a further study for WLFs. In fact, the relationship between the 360 nm emission and the emission at $>$400 nm, i.e. the true WL, is worthwhile to be explored.

In the present study, the WL emission at 360 nm reaches its peak around the same time as the temporal derivative of the SXR 1\,--\,8 \AA\ flux as well as the HXR 20\,--\,50 keV emission in most WLFs, i.e. exhibiting the Neupert effect. This suggests that the 360 nm emission can be closely related to a nonthermal electron beam heating, i.e. having a nonthermal origin. It should be noted that there are two exceptional WLFs in our data set, whose emissions at 360 nm have a later peak time than the SXR 1\,--\,8 \AA\ flux. This kind of emission appearing in the decay phase of the flare (or gradual-phase emissions) might be produced by a thermal process (say, thermal conduction or SXR backwarming) rather than a nonthermal electron-beam heating that usually occurs in the rise phase. These are worthy of a detailed study in the future. 

In our study, the WL emission at 360 nm exhibits a center-to-limb variation, which is manifested as a larger enhancement plus more WLFs identified near the limb. This is consistent with previous studies on WL emission both in observations and numerical simulations. A statistical work on 86 WLFs showed that the WLFs prefer to occur at the solar limb (\citealp{1993SoPh..144..169N}). Some numerical simulations revealed that the electron-beam heating leads to a larger enhancement in the continuum near the solar limb than at the disk center (e.g. \citealp{2007ASPC..368..417D,2010ApJ...711..185C}). The authors gave an explanation that the formation height of continuum emission is higher in limb flares and it is easier to heat. Our study provides an observational evidence for the center-to-limb effect of the 360 nm emission from a statistical analysis. Furthermore, we find that the center-to-limb variation is more obvious for larger WLFs.

At the time of writing, WST is beginning to provide WL images with a very rapid cadence of one or two seconds in a burst mode. This can bring some new sights on WLFs at 360 nm. Firstly, a high temporal resolution can provide precise start, peak, and end times of WL emission as well as their relationships with the corresponding times of SXR and HXR emissions, which helps understand the physical origin of the 360 nm emission. A rapid cadence of WL images might also make it possible to detect WL waves at 360 nm accompanied by WLFs (like sunquakes observed from HMI; \citealp{2023ApJ...943L...6W}). This would be very helpful to explore the conditions in the WL emission source. As more and more WLFs are collected from WST observations, it is worthwhile doing a comprehensive statistical work combining with the data from some other instruments such as HMI, ONSET, and the \textit{Chinese H$\alpha$ Solar Explorer} (CHASE: \citealp{2022SCPMA..6589602L}). On the other hand, radiative-hydrodynamic simulations are needed to help explain the observations in terms of formation height and contribution function of the WL emission. All these should improve our understanding of WLFs at 360 nm.

\begin{acks}
The authors thank the reviewer very much for the valuable, as well as, detailed suggestions and comments that helped to improve the manuscript greatly. The ASO-S mission is supported by the Strategic Priority Research Program on Space Science, Chinese Academy of Sciences. SDO is a mission for NASA’s Living With a Star program. 
\end{acks}

\begin{authorcontribution}
Z.C. Jing analyzed the data and wrote the first manuscript. Y. Li proposed the initial idea, helped the data analysis, and revised the manuscript. J. Tian and D.C. Song provided suggestions on the data analysis. W.Q. Gan is PI of ASO-S. H. Li and L. Feng are PI and Co-PI of LST, respectively. Y. Huang, Y.P. Li, J.W. Li, J.C. Xue, J. Zhao, L. Lu, B.L. Ying, P. Zhang, Y.N. Su, Q.M. Zhang, D. Li, Y.Y. Ge, S.T. Li, Q. Li, G. Li, X.F. Liu, G.L. Shi, J.H. Shan, Z.Y. Tian, and Y. Zhou contributed on the pipeline and release of WST data. Y. Su and W. Chen provided the HXI data. All authors reviewed the manuscript.
\end{authorcontribution}

\begin{fundinginformation}
The authors are supported by the Strategic Priority Research Program of the Chinese Academy of Sciences (Grant No. XDB0560000), the National Natural Science Foundation of China (Grant Nos. 12273115, 12233012, and 11921003), and the Ministry of Science and Technology of China (Grant No. 2022YFF0503004).
\end{fundinginformation}

\begin{dataavailability}
Data of ASO-S before 1 April 2023 are under test and not publicly available, but they are available from the corresponding author on reasonable request. Data of ASO-S after 1 April 2023 are publicly available and can be downloaded from the official website of ASO-S at \urlurl{aso-s.pmo.ac.cn/sodc/dataArchive.jsp}. Data of SDO used in this work are publicly available at \urlurl{jsoc.stanford.edu}.
\end{dataavailability}

\begin{ethics}
\begin{conflict}
The authors declare no competing interests.
\end{conflict}
\end{ethics}

\bibliographystyle{spr-mp-sola}
\bibliography{refer}  

\newpage

\begin{table}[htb]
\caption{The 39 WLFs analyzed in detail.}
\label{tab:all}
\tiny
\setlength{\tabcolsep}{1.0mm}{
\begin{tabular}{cccccccccccccccc}                               
\hline                   
 Observation & Flare & GOES & GOES & HXR & WL & WL & WL & WL & WL & WL & WL & WL \\
  Date & Location\tabnote{The flare location is derived from the GOES flare catalog.} & Class & Peak & Peak\tabnote{The HXR peak means the peak time of the HXI 20\,--\,50 keV emission. Note that some WLFs have no available data from HXI, which are tagged as ``-''.} & Start & Peak & End & $\tau$\tabnote{There are three flares whose WL duration is less than two minutes. This is because their WL brightenings only appear in one time frame.} & $r_\mathrm{p}$ & $r_\mathrm{m}$ & $S$\tabnote{The uncertainty comes from the maximum difference between the WL brightening locations and the flare location in the process of correcting project effect.} & $F$\tabnote{The uncertainty of total flux consists of two parts, one from the average photometric fluctuation of 1.5\% and the other from the uncertainty of the radiometric calibration factor that is obtained via a comparison between the observed counts of WST and standard irradiance at 360 nm from ASTM G173-03.}\\
     &  &  & [UT] & [UT] & [UT] & [UT] & [UT] & [minute] & [\%] & [\%] & [arcsec$^2$] & [erg]\\
  \hline

07.11.2022 & N12E56 & M5.2 & 00:11 & 00:05 & 00:07 & 00:07 & 00:13 & 6.0 & 16.9 & 9.94 & 404$\pm$20 & (5.33$\pm$0.09)$\times10^{26}$\\
14.12.2022 & S21W48 & M4.5 & 22:06 & - & 22:07 & 22:15 & 22:39 & 32 & 24.6 & 13.8 & 194$\pm$6.4 & (1.71$\pm$0.03)$\times10^{29}$\\
16.12.2022 & S21W64 & M3.5 & 02:01 & - & 01:40 & 01:54 & 01:48 & 8.0 & 25.8 & 15.2 & 225$\pm$10 & (4.22$\pm$0.07)$\times10^{28}$\\
16.12.2022 & S20W68 & M4.0 & 10:19 & 10:18 & 10:16 & 10:20 & 10:36 & 20 & 28.7 & 17.0 & 233$\pm$30 &(5.40$\pm$0.08)$\times10^{28}$\\
30.12.2022 & N20E07 & M3.7 & 19:38 & 19:37 & 19:36 & 19:36 & 19:36 & $<$2.0 & 10.9 & 9.47 & 13.0$\pm$0.65 &(8.04$\pm$0.13)$\times10^{26}$\\
09.01.2023 & N24E79 & X1.9 & 18:50 & - & 18:46 & 18:50 & 19:02 & 16 & 101 & 31.9 &  (2.62$\pm$0.33)$\times10^{3}$ &(1.91$\pm$0.03) $\times10^{30}$\\
10.01.2023 & S13E71 & M5.1 & 00:16 & 00:14 & 00:14 & 00:14 & 00:24 & 10 & 34.7 & 18.1 & (1.65$\pm$0.32)$\times10^{3}$ &(2.62$\pm$0.06)$\times10^{29}$\\
10.01.2023 & S12E68 & M1.0 & 02:16 & 02:13 & 02:14 & 02:16 & 02:18 & 4.0 & 23.0 & 12.8 & 163$\pm$19 &(1.95$\pm$0.03)$\times10^{28}$\\
10.01.2023 & S13E59 & M1.2 & 17:48 & 17:47 & 17:48 & 17:48 & 17:48 & $<$2.0 & 30.6 & 20.3 & 68.9$\pm$6.0 &(5.92$\pm$0.10)$\times10^{27}$\\
10.01.2023 & N22E62 & X1.0 & 22:47 & - & 22:46 & 22:48 & 23:00 & 14 & 83.1 & 31.8 & 720$\pm$92 &(3.28$\pm$0.06)$\times10^{29}$\\
11.01.2023 & N25E56 & M2.4 & 00:59 & - & 00:52 & 00:54 & 01:10 & 18 & 16.0 & 10.4 & 176$\pm$19& (1.41$\pm$0.02)$\times10^{28}$\\
11.01.2023 & S13E53 & M5.6 & 01:56 & 01:53 & 01:54 & 01:54 & 01:58 & 4.0 & 34.2 & 17.6 & 410$\pm$21 &(4.97$\pm$0.08)$\times10^{28}$\\
11.01.2023 & N22E56 & M3.1 & 08:33 & - & 08:32 & 08:34 & 08:38 & 6.0 & 71.9 & 24.8 & 276$\pm$6.3 &(5.42$\pm$0.10)$\times10^{28}$\\
12.01.2023 & N23E49 & M1.6 & 06:50 & - & 06:56 & 07:00 & 07:06 & 10 & 21.1 & 12.5 & 127$\pm$7.2 &(1.03$\pm$0.02)$\times10^{28}$\\
13.01.2023 & S11W83 & M4.0 & 10:15 & - & 10:13 & 10:13 & 10:15 & 2.0 & 33.1 & 17.4 & (1.59$\pm$0.33)$\times10^{3}$ &(3.18$\pm$0.05)$\times10^{29}$\\
07.02.2023 & N28E02 & M3.8 & 22:58 & - & 22:53 & 22:55 & 23:03 & 10 & 61.0 & 13.3 &  65.4$\pm$1.9 &(6.99$\pm$0.11)$\times10^{27}$\\
07.02.2023 & N30E03 & M6.4 & 23:07 & 23:06 & 23:07 & 23:07 & 23:09 & 2.0 & 55.5 & 22.8 & 230$\pm$3.5 & (2.53$\pm$0.04)$\times10^{28}$\\
08.02.2023 & N30E01 & M2.1 & 02:53 & 02:53 & 02:53 & 02:53 & 02:53 & $<$2.0 & 31.6 & 16.0 & 67.0$\pm$1.2 &(5.06$\pm$0.08)$\times10^{27}$\\
09.02.2023 & S09E69 & M1.1 & 07:17 & 07:17 & 07:17 & 07:17 & 07:19 & 2.0 & 23.2 & 15.0 & 273$\pm$21 &(2.58$\pm$0.04)$\times10^{28}$\\
10.02.2023 & N30W23 & M3.7 & 03:03 & - & 02:57 & 02:59 & 03:07 & 10 & 22.5 & 13.8 & 89.1$\pm$3.8 &(6.26$\pm$0.10)$\times10^{27}$\\
10.02.2023 & N31W27 & M1.4 & 08:05 & 08:05 & 08:05 & 08:05 & 08:07 & 2.0 & 26.0 & 14.8 & 82.7$\pm$3.7 & (7.95$\pm$0.12)$\times10^{27}$\\
11.02.2023 & N07W69 & M2.2 & 08:08 & 08:06 & 08:03 & 08:05 & 08:07 & 4.0 & 16.9 & 11.9 & 18.3$\pm$0.41 & (3.07$\pm$0.05)$\times10^{27}$\\
11.02.2023 & S10E39 & X1.1 & 15:48 & 15:46 & 15:41 & 15:47 & 16:01 & 20 & 51.8 & 23.9 & 841$\pm$15 &(2.71$\pm$0.05)$\times10^{29}$\\
12.02.2023 & S10E29 & M3.1 & 08:48 & 08:44 & 08:45 & 08:45 & 08:51 & 6.0 & 25.8 & 16.1 & 98.8$\pm$1.8 & (1.36$\pm$0.02)$\times10^{28}$\\
17.02.2023 & N25E67 & X2.3 & 20:16 & 20:02 & 19:58 & 20:02 & 20:42 & 44 & 61.6 & 24.4 & (1.77$\pm$0.25)$\times10^{3}$ &(1.45$\pm$0.02)$\times10^{30}$\\
03.03.2023 & N21W76 & X2.1 & 17:52 & - & 17:50 & 17:50 & 18:06 & 16 & 203 & 62.1 & (2.32$\pm$0.43)$\times10^{3}$&(5.83$\pm$0.10)$\times10^{29}$\\
06.03.2023 & N18W63 & M5.8 & 02:28 & 02:21 & 02:20 & 02:22 & 02:32 & 12 & 32.3 & 13.1 & 222$\pm$41 & (1.94$\pm$0.03)$\times10^{29}$\\
29.03.2023 & S21W59 & X1.2 & 02:33 & - & 02:30 & 02:32 & 02:40 & 10 & 45.6 & 17.3 & 748$\pm$126 &(3.36$\pm$0.05)$\times10^{29}$\\
30.03.2023 & S22W77 & M5.4 & 07:37 & - & 07:33 & 07:35 & 07:37 & 4.0 & 25.9 & 16.6 & (1.08$\pm$0.11)$\times10^{3}$& (1.02$\pm$0.05)$\times10^{29}$\\
27.04.2023 & S22E05 & M1.8 & 11:14 & 11:12 & 11:13 & 11:13 & 11:19 & 6.0 & 29.5 & 17.4 & 74.4$\pm$0.81 & (2.19$\pm$0.04)$\times10^{28}$\\
03.05.2023 & N13E43 & M4.3 & 09:27 & - & 09:23 & 09:25 & 09:31 & 8.4 & 36.0 & 20.0 & 141$\pm$2.1 &(2.03$\pm$0.04)$\times10^{28}$\\
03.05.2023 & N12E42 & M3.1 & 10:14 & 10:12 & 10:09 & 10:11 & 10:11 & 1.7 & 16.5 & 12.6 &  32.4$\pm$0.30 & (5.82$\pm$0.12)$\times10^{26}$\\
03.05.2023 & N12E42 & M7.2 & 10:45 & 10:40 & 10:44 & 10:44 & 10:52 & 7.8 & 67.9 & 28.7 & 234$\pm$2.0 &(6.25$\pm$0.12)$\times10^{28}$\\
03.05.2023 & N12E41 & M1.7 & 12:35 & 12:34 & 12:30 & 12:35 & 12:44 & 15 & 30.2 & 21.9 & 76.3$\pm$0.64 &(1.24$\pm$0.02)$\times10^{28}$\\
03.05.2023 & N12E40 & M2.2 & 13:50 & 13:47 & 13:47 & 13:47 & 13:53 & 6.6 & 38.7 & 20.6 & 125$\pm$1.0 &(2.55$\pm$0.05)$\times10^{28}$\\
09.05.2023 & N13W26 & M6.5 & 03:54 & 03:51 & 03:47 & 03:53 & 04:11 & 24 & 35.9 & 20.0 & 111$\pm$6.1 &(4.86$\pm$0.08)$\times10^{28}$\\
18.05.2023 & N17E77 & M3.9 & 20:23 & 20:22 & 20:22 & 20:22 & 20:28 & 6.0 & 43.3 & 22.9 & 417$\pm$64 &(1.53$\pm$0.03)$\times10^{29}$\\
19.05.2023 & N17E79 & M5.4 & 00:48 & 00:44 & 00:44 & 00:44 & 00:50 & 6.0 & 32.7 & 22.1 & 298$\pm$45 &(7.26$\pm$0.13)$\times10^{28}$\\
20.05.2023 & N17E55 & M5.7 & 15:00 & 14:59 & 14:52 & 14:58 & 15:14 & 22 & 56.7 & 25.8 & 396$\pm$7.0 &(1.33$\pm$0.03)$\times10^{29}$\\
  \hline
\end{tabular}
}
\end{table}

\end{article} 
\end{document}